\documentclass[conference]{IEEEtran}

\usepackage[noadjust]{cite}
\usepackage{amsmath,amssymb,amsfonts}
\usepackage{algorithmic}
\usepackage{graphicx}
\usepackage{textcomp}
\usepackage{caption}
\usepackage{subcaption}
\usepackage{xcolor}
\usepackage[hyphens]{url}
\def\BibTeX{{\rm B\kern-.05em{\sc i\kern-.025em b}\kern-.08em
    T\kern-.1667em\lower.7ex\hbox{E}\kern-.125emX}}
\usepackage{autobreak}
\newcommand{\CheckUpdate}{\mathsf{CheckUpdate}}

\newcommand{\Witness}{\mathsf{Witness}}
\newcommand{\Belongs}{\mathsf{Belongs}}
\newcommand{\Update}{\mathsf{Update_{op}}}

\newcommand{\Setup}{\mathsf{Setup}}

\newcommand{\totalSupply}{\mathbf{totalSupply}}
\newcommand{\balanceOf}{\mathbf{balanceOf}}
\newcommand{\allowance}{\mathbf{allowance}}
\newcommand{\transfer}{\mathbf{transfer}}
\newcommand{\approve}{\mathbf{approve}}
\newcommand{\transferFrom}{\mathbf{transferFrom}}

\begin{document}

% \title{Revisiting the ERC20 Token Standard and Ethereum's Storage Cost Model
% }
\title{An Alternative Paradigm for Developing and Pricing Storage on Smart
Contract Platforms
}

\author{\IEEEauthorblockN{Christos Patsonakis}
\IEEEauthorblockA{\textit{Department of Informatics and Telecommunications} \\
\textit{University of Athens}\\
c.patswnakis@di.uoa.gr}
\and
\IEEEauthorblockN{Mema Roussopoulos}
\IEEEauthorblockA{\textit{Department of Informatics and Telecommunications} \\
\textit{University of Athens}\\
mema@di.uoa.gr}
}
\maketitle

\begin{abstract}

% Bitcoin and its underlying blockchain technology laid the foundation for
% the introduction of modern platforms that allow the development of 
% smart contracts. 
Smart contract platforms, the most notable of which is probably Ethereum, facilitate
the development of important and diverse distributed applications (e.g., naming services and fungible tokens) in a simple manner. 
This simplicity stems from the inherent utility of employing the state of
smart contracts to store, query and verify the validity of application
data. In Ethereum, data storage incurs an underpriced,
non-recurring, predefined fee. Furthermore, as there is no incentive 
for freeing or minimizing the state of smart contracts, Ethereum is faced with a tragedy of the commons problem with 
regards to its monotonically increasing state. This issue, if left
unchecked, may lead to centralization and directly impact Ethereum's security
and longevity. 

In this work, we introduce an alternative paradigm for developing smart contracts
in which their state is of constant size and facilitates the verification of
application data that are stored to and queried from an external, potentially
unreliable, storage network. This approach is relevant for a wide range of applications, such as any key-value store. We evaluate our approach by adapting
the most widely deployed standard for fungible tokens, i.e., the ERC20
token standard. We show that Ethereum's current cost model penalizes our
approach, even though it minimizes the overhead to Ethereum's state and
aligns well with Ethereum's future. We address Ethereum's monotonically
increasing state in a two-fold manner. First, we introduce recurring
fees that are proportional to the state of smart
contracts and adjustable by the miners that maintain the network. Second,
we propose a scheme where the cost of storage-related operations reflects
the effort that miners have to expend to execute them. Lastly, we show
that under such a pricing scheme that encourages economy in the state consumed by
smart contracts, our ERC20 token adaptation reduces the incurred
transaction fees by up to an order of magnitude.

\end{abstract}

\section{Introduction}
\label{sec:introduction}

% Bitcoin (\cite{bitcoin}) revolutionized the world of digital payments by allowing untrusted
% entities to transact securely without relying on trusted, third parties. Its
% operation is based on a distributed network of peers with open membership
% that maintains a highly replicated, auditable, append-only log of transactions,
% which is commonly referred to as a blockchain. Nodes are incentivized to
% participate in the network via digital currency rewards. A second generation of
% blockchains
% pushes the limit of potential applications further from the exchange of digital
% currency. They allow the development of smart contracts (\cite{smartcontracts}),
% i.e., digital agents that encode, execute and enforce arbitrary agreements
% between untrusted parties.
% 
% A smart contract platform models a stateful, global computer where multiple applications
% can coexist and interact with each other. These platforms provide the means of
% developing diverse and important distributed applications (dApps) in a simple
% manner that, prior to their introduction, was challenging to implement.

Bitcoin (\cite{bitcoin}) revolutionized the world of digital payments by allowing untrusted
entities to transact securely without relying on trusted, third parties. Its
operation is based on a distributed network of peers with open membership
that maintains a highly replicated, auditable, append-only log of transactions,
which is commonly referred to as a blockchain. A second generation of
blockchains
allows the development of smart contracts (\cite{smartcontracts}),
i.e., digital agents that encode, execute and enforce arbitrary agreements. Smart contract platforms provide the means of
developing diverse and important distributed applications (dApps) in a simple
manner that, prior to their introduction, was challenging to implement.

Ethereum (\cite{ethyellowpaper}) is probably the most notable smart contract platform. Its live chain 
features dApps that implement naming services (\cite{ethens}), multisignature
wallets (\cite{ethmultisig}), a large variety of fungible tokens 
(\cite{erc20tokencontracts}) and even crypto-collectibles
(\cite{cryptokitties}), all in just a few lines of code.
The simplicity of developing dApps on top of these platforms stems from the inherent
utility of employing the state of smart contracts to store, query and verify
the validity of application data. For instance, all implementations of the most
widely deployed standard for fungible tokens, i.e., the ERC20 token standard
(\cite{erc20}),
store each account's token balance on the contract's state. 

Today, Ethereum's cost model does not adequately take into account
the amount of storage consumed by smart contracts.  This is problematic
for several reasons.
First, in Ethereum,
storing data on the state of smart contracts requires paying  
\emph{one, non-recurring fee} at the time the
data is stored. Thus, regardless of the amount of state that they 
consume, contracts have zero maintenance costs and can be part of 
Ethereum's state
\textbf{forever}. Second, storage-related operations are 
underpriced, as stated by Ethereum's creator, Vitalik Buterin, in one of his recent talks (\cite{transactionfeeeco}).  
These two factors facilitate contracts
that gain utility from storing small amounts of data per user and have low computational complexity, such
as ERC20 tokens.  
As a result, such contracts have very low transaction fees for 
their operations. 
Third and most importantly,
Ethereum's state must be maintained by all full nodes, yet
there is no incentive mechanism in place for freeing storage.  If left
unchecked, this can have serious consequences.
It will diminish the mining population
as proportionally fewer and fewer miners will be able to contribute 
to the network.
This will lead to centralization and may prohibit new nodes from
joining and syncing to the
network. This will have a direct impact on Ethereum's 
security and, utlimately, its longevity.
%XXX Maybe we could say something about
%the bandwidth of miners here?? Not sure if it fits or how we can make it stand XXX.

In this work, we introduce an alternative paradigm for developing dApps on top
of smart contract platforms by decoupling the issue of storage from verifying
the validity of data. The former is handled by an external, potentially 
unreliable, storage network that allows efficient access
to the application's data. To verify the validity of data obtained from the
storage network, we maintain cryptographic accumulators in the smart contract's
state. 
These are data structures that provide a constant-sized representation of
a set of elements and allow for verifiable (non) membership proofs. 
To evaluate
our approach, we present a case study of an accumulator-based 
implementation of the ERC20 token standard. 
We choose this standard because it is the most widely deployed token standard for 
fungible tokens, numbering over 130,000 compliant contracts on Ethereum's live
chain (\cite{erc20tokencontracts}). 
Via minor modifications, our construction can be modified
to fit other, upcoming standards, such as the ERC721 standard (\cite{erc721}) for non-fungible
tokens. However, we stress that our approach can be adapted to any application that requires a verifiable representation of its application data, e.g., naming services, voting systems or any kind of key-value store. 

By requiring only minimal (constant-sized) state to be stored
in the contract, our accumulator-based approach promotes diversity, 
scalability, and security of the Ethereum network.  Yet, we 
show that under Ethereum's current cost model, this accumulator-based approach
is penalized for the security properties it provides;  it is much more
(almost prohibitively) 
costly than the approach of
storing each account's token balance in the contract state.
This illustrates one of Ethereum's main incentive misalignments. To
address this, we revisit Ethereum's storage cost model and propose 
modifications that: 1) price storage-related operations based on the effort that 
miners have to expend to execute them, 2) ensure that contracts pay recurring
fees proportionate to the amount of storage they consume and the system's overall
capacity and, 3) free space consumed by unused/stale contracts. We show
that under such a pricing scheme, our accumulator-based ERC20 token 
construction reduces the incurred transaction fees by up to an order of
magnitude. With these modifications, we hope the Ethereum developer community will
be encouraged to exercise economy in the state consumed by the smart contracts
they develop.

\section{Ethereum}
\label{sec:ethereum}

Ethereum is a blockchain-based, 32-byte word, global computer that allows the development of smart
contracts, i.e., stateful agents that ``live'' in the blockchain and can execute arbitrary state transition functions. Smart contract code is written in
a high-level, Turing-complete programming language
(e.g., Solidity~\cite{solidity}), which is then compiled-down to Ethereum Virtual Machine (EVM)
initialization code. Contracts are deployed by wrapping their initialization code in a transaction, signing
it and broadcasting it to the network. Users can interact with smart
contracts by broadcasting appropriately formatted transactions.
Smart contracts are ``passive'' entities that, as a result of a user's
transaction, can issue \emph{message calls}, i.e., call functions of other
contracts.
Ethereum's cryptocurrency is called \emph{ether} and serves as a means to incentivize participants
(miners) to engage in the protocol. 
Transactions fees are measured in a unit called
\emph{gas} and are a function of the byte size and the complexity of the code invoked by transactions
(if any). Each transaction byte and EVM operation
costs some predefined amount of gas (\cite{ethyellowpaper}). Transactions specify a \emph{gas
price}, which converts ether to gas and influences the incentive of miners to include it in
their next block. A transaction that consumes $g_{cost}$ gas and specifies a
gas price of $g_{price}$ will cost $E = g_{cost} \times g_{price}$ units of ether.
Lastly, transactions and message calls, specify an upper bound on the amount of gas that they
can consume. This protects miners from, e.g., getting stuck in an infinite loop, an issue that
stems from Ethereum's Turing-completeness.

\section{Hash Tree Universal Accumulator}
\label{sec:hashaccumulator}

Cryptographic accumulators provide a constant representation of a set of elements and
allow for verifiable membership queries. Universal accumulators also allow for
verifiable non membership queries. Proving statements (e.g., element membership) is
facilitated via values that are referred to as \emph{witnesses}. Informally, the
security property of accumulators states that an adversary is unable to generate a
valid witness value for a false statement, except with negligible probability. For
instance, an adversary is not able to generate a valid membership witness for an
element $x$ that is not part of the accumulated set of elements $X$. It is common
to refer to the party that maintains and manages the accumulator as the
\emph{accumulator manager}. In our accumulator-based ERC20 token, this role is played
by the smart contract.

In the following, we provide a high level description of the hash tree,
universal accumulator of Camacho et al.~\cite{hashacc}, whose security is based on
collision-resistant hash functions. This accumulator employs
a public data structure $m=(T,X)$ (referred to as \emph{memory}),
where $X = \{x_1,...,x_n\}$ is the set of accumulated elements and $T$ is a
binary, balanced hash tree. The accumulator's value (denoted as $Acc$) is the hash of
$T$'s root node. Camacho et al.~\cite{hashacc} model their accumulator as a tuple of the
following algorithms:

\begin{itemize}
 \item $\Setup\mathsf{(k)}:$ On input the security parameter $k \in \mathbb{N}$, it outputs the
 accumulator's initial value $Acc_0 \in \{0,1\}^k$, which corresponds to the set
 $X = \emptyset$, and an initialized memory $m_0$. 
 \item $\Witness\mathsf{(Acc,m,x)}:$ This algorithm outputs a membership or a non membership witness $W$, if $x \in X$ or if $x \notin X$, respectively. 
 \item $\Belongs\mathsf{(Acc,x,W)}:$ This algorithm outputs $1$, if $W$ is a valid witness for 
 $x \in X$, $0$, if $W$ is  a valid witness for $x \notin X$, or $\perp$ otherwise.
 \item $\Update\mathsf{(Acc_{\mathsf{before}},m_{\mathsf{before}},x)}:$ This algorithm updates the accumulator's
 value by either adding ($\mathsf{op = add}$) or removing ($\mathsf{op = del}$) the element $x$ to/from
 the accumulated set $X$. It outputs the updated values of the accumulator
 ($\mathsf{Acc_{after}}$) and its memory ($\mathsf{m_{after}}$), as well as, an update witness $\mathsf{W_{op}}$.
 \item $\CheckUpdate\mathsf{(Acc_{\mathsf{before}},Acc_{\mathsf{after}},x,W_{\mathsf{op}})}:$ This algorithm outputs $1$,
 if $W_{op}$ is a valid witness for an update operation ($\mathsf{op \in \{add,del\}}$)
 pertaining element $x$, which updated the accumulator's value from $\mathsf{Acc_{before}}$ to 
 $\mathsf{Acc_{after}}$, otherwise, it outputs $0$.
\end{itemize}

% In the following, we highlight some important properties of this accumulator. First,
This accumulator is strong, i.e., it does not require a trusted setup nor a trusted
accumulator manager. It allows for updates (additions and deletions) that
can be performed without having access to secret information and are 
\emph{publicly verifiable}. The latter is accomplished via the $\CheckUpdate$ algorithm
which, on input a witness returned by $\Update$ and the accumulator's values before
and after the update, outputs 1, if the update was performed honestly, and 0,
otherwise. In this accumulator, (non) membership and update witnesses are hash path(s)
starting from some node(s) in $T$ (not necessarily leaf node(s)) that lead all the way
up to the root node. Thus, their size is $\mathcal{O}(k \log_2(n))$, where $n = |X|$.

\section{Accumulator-based ERC20 Token}
\label{sec:accerc20}

The ERC20 token standard (\cite{erc20}) describes the functions and events that
facilitate the exchange of arbitrary crypto-assets.
% At the time of this writing, it is the most
% widely deployed token standard, numbering over 130,000 compliant smart contracts
% (\cite{erc20tokencontracts}) on the Ethereum live chain.
Each token holder's account
is associated with an Ethereum $\mathbf{address}$ data type. The token balance of
each account is commonly represented as a $\mathbf{uint}$ data type, i.e., an unsigned integer. The ERC20 token interface is comprised of the following functions:

\begin{enumerate}
 \item $\totalSupply\mathbf{()\!:}$ Outputs the total supply of tokens
 accross all accounts.
 \item $\balanceOf\mathbf{(address\; owner)\!:}$ Outputs the token balance of the input account.
 \item $\approve\mathbf{(address\;spender,uint\;tokens)\!:}$ The account that issues the call
 (transaction) to this function authorizes the ``$\mathbf{spender}$'' account to transfer
 the specified number of $\mathbf{tokens}$ from her account.
 \item $\allowance\mathbf{(address\; owner,address\;spender):}$ Outputs the
 number of tokens that the spender's account is $\approve$'d to transfer from the
 owner's account.
 \item $\transfer\mathbf{(address\;to,uint\;tokens)\!:}$ The account that issues the call
 (transaction) to this function transfers the specified number of tokens to the
 ``$\mathbf{to}$'' account.
 \item $\begin{aligned}[t]
    &\transferFrom\mathbf{(address\;from,address\;to,uint}\\[-1mm]
    &\mathbf{tokens):}
    \end{aligned}$ 
    \\\\[-1.25\baselineskip]
    \hspace*{14mm} Transfers the specified number of
    $\mathbf{tokens}$ from  account ''$\mathbf{from}$`` to the
    $\approve$'d account ''$\mathbf{to}$``.

%  \begin{align*}\transferFrom\mathbf{(address\;from,address\;to,uint\;to-}\\ &&\text{kens):} \end{align*}

%  
%  \text{This
%  function transfers the specified number of }\mathbf{tokens}\text{ from the
%  ``}\mathbf{from}\text{'' account to the }\approve\text{'d account ``}\mathbf{to}\text{''.}\end{aligned}$
\end{enumerate}

To facilitate the aforementioned functionality, ERC20 compliant smart contracts 
store two mappings in their state: 1) \emph{balances},
which maps account addresses to token balances and, 2) \emph{allowed}, which maps
account addresses to another mapping where, the latter, maintains the balance that
each $\approve$'d account is allowed to transfer from the token owner's account.

% In this work, we propose a different paradigm for developing smart contracts
% in which we decouple querying and storage of data from their verification.
% Querying and storage of data is handled by an external,
% potentially unreliable, storage network,
% which allows
% for a more compact retrieval of the application's state. To verify the validity of
% the information retrieved by the storage network, we store cryptographic accumulators
% at the smart contract's state.

We now illustrate how we employ the hash tree, universal accumulator of
Camacho et al.~\cite{hashacc} (Section~\ref{sec:hashaccumulator}), to realize an accumulator-based ERC20 token. The
core idea is to replace each aforementioned mapping with one accumulator. We
replace the \emph{balances} mapping with an accumulator, \emph{balancesAcc}, that
accumulates (owner,tokens) tuples and allows clients to infer each account's token balance. For the \emph{allowed} mapping, which is a ``double'' mapping, we
need two accumulators. The first accumulator, \emph{allowedAddressesAcc}, accumulates
(owner,spender) tuples and allows clients to infer the accounts that token
owners have $\approve$'d. The second accumulator, \emph{allowedBalancesAcc},
accumulates (spender,tokens) tuples and allows clients to infer the token balance
that $\approve$'d accounts are allowed to transfer from the owner's account. 
Thus, we have a constant-sized and verifiable representation of account balances
and allowances.

% Due to space limitations, we are unable to present a concrete specification of the
% storage network and how clients interact with it as it would require a
% more in depth description of the hash tree accumulator of Camacho et al.
% \cite{hashacc}. However, we briefly discuss some important points and
% refer the interested reader for more details and the security proof of
% our construction to the extended paper.

Our
design's security depends solely on that of the smart contract platform (Ethereum in our case) and the accumulator scheme. This
allows us to employ a variety of primitives to realize the storage network,
whose concrete specification we leave as future work. For instance,
even centralized cloud storage services are a viable option. However, we believe that the best approach is a distributed file
storage system, especially one that has ``bridges'' with the Ethereum
network. Some notable examples are Swarm (\cite{swarm}), Storj (\cite{storj}) and IPFS
(\cite{ipfs}). The storage network's state is assumed
to be
comprised of the \emph{memory} data structure (see Section \ref{sec:hashaccumulator})
of each of the aforementioned accumulators. As we show below, 
the interaction with accumulator-based ERC20 smart contracts 
requires the construction of (non) membership and update witnesses by the
clients which, subsequently, are subject to verification by the smart contract.
Clients construct these witnesses by interacting with the storage network. We stress that clients do not need
to download the entire \emph{memory} of accumulators to construct these witnesses. The data that needs to be
transmitted from storage nodes to clients are hash paths from the appropriate
accumulators' hash trees, i.e., they are of logarithmic complexity. Thus, from hereon
in, we assume that clients can efficiently construct the witness values that are
required to realize the ERC20 token interface.

Accumulator-based ERC20 token smart contracts cannot implement the $\balanceOf$ and
$\allowance$ functions since they do not store account balances and allowances in
their state. Instead, clients are able to infer the information obtained by these
functions by interacting with the storage network. To infer the balance $y$ of account $x$, clients construct and verify a membership witness that the 
tuple $(x,y)$ is accumulated in \emph{balancesAcc}. To infer the allowance
$z$ of a spender's account $x_2$ from an owner's account $x_1$, clients construct and verify two membership witnesses. First, a membership
witness that the tuple $(x_1,x_2)$ is accumulated in
\emph{allowedAddressesAcc}, which proves that the token owner $x_1$ has allowed the spender account $x_2$ to transfer some tokens from her account. Second, a membership witness that the tuple $(x_2,z)$ is accumulated in
\emph{allowedBalancesAcc}, which proves the number of tokens the spender
is allowed to transfer from the token owner's account.

An account $x_1$ with balance $y_1$ that wishes to $\transfer$ $z$ tokens
($y_1 \geq z$) to an account $x_2$ with balance $y_2$ produces the following proofs.
First, a membership witness for the tuple $(x_1,y_1)$ in \emph{balancesAcc}, which
proves the owner's account balance. Second, a membership witness for the tuple
$(x_2,y_2)$, which proves the balance of the destination account. Third,
an update witness for the deletion of the tuple $(x_1,y_1)$ from \emph{balancesAcc}.
Fourth, an update witness for the deletion of the tuple $(x_2,y_2)$ from
\emph{balancesAcc}. Fifth, an update witness for the addition of the tuple
$(x_1,y_1-z)$ to \emph{balancesAcc}. Sixth, an update witness for the addition of
the tuple $(x_2,y_2+z)$ to \emph{balancesAcc}. Notice that the sequence
of the involved updates reflects the transfer of $z$ tokens from $x_1$ to  $x_2$.

Due to space limitations, we are unable to describe how we realize the
$\approve$ and $\transferFrom$ operations. To provide insight with
regards to their complexity, we mention the proofs that are involved in
each operation. The $\approve$ operation involves two update witnesses and
either one non membership, or, one membership witness, depending on 
whether the token owner approves the spender's account for the first time 
or not, respectively. The $\transferFrom$ operation involves four membership witnesses
and six update witnesses. Thus, the $\transferFrom$
is the most expensive operation, followed by $\transfer$ and, lastly,
by $\approve$.

\section{Evaluation}
\label{sec:evaluation}

In this section, we evaluate our accumulator-based ERC20 token construction.
We ran our experiments on a private blockchain that is maintained by a single
mining node. We use the latest, stable version of \emph{geth} (v1.8.17,
\cite{gethrelease}), Ethereum's official client, that was available at the
time of this writing. We conducted our experiments via the \emph{truffle}
suite (v4.1.13, \cite{truffle}) that employs \emph{solc-js} (v0.4.24,
\cite{solc}) to compile smart contracts with optimizations enabled.

\begin{figure}
%\centering
        \includegraphics[width=.7\linewidth,angle=270]{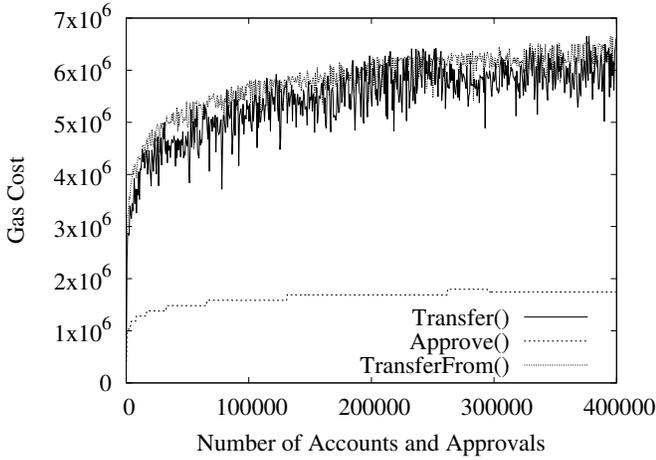} 
\caption{Gas cost versus of the
$\transfer$, $\approve$ and $\transferFrom$ operations of our 
accumulator-based ERC20 token construction for up to a total of 400,000
accounts and 400,000 approvals.}
\label{fig:accerc20}
\end{figure}

Figure~\ref{fig:accerc20} illustrates the gas cost of the $\transfer$, 
$\approve$ and $\transferFrom$ operations of our accumulator-based ERC20 token for up to a total of 400,000
accounts and 400,000 approvals. Results illustrate that 
transaction gas costs scale logarithmically, which is expected (same holds for
the $\approve$ operation). Recall that all involved proofs are hash path(s)
starting from some node(s) (not necessarily leaf node(s)) in the accumulator's
tree. Thus, their size and verification cost varies based
on the position of those nodes in the tree. Our construction's operations consume a large portion
of the block's limit which is, currently, about 8 million gas
(\cite{gaslimit}). In the following, we discuss a
series of improvements that will diminish the cost of our construction's operations.

% To illustrate the overhead of our
% accumulator-based ERC20 token construction, we implemented a bare-bones
% ERC20 token (\cite{sampleerc20}) and repeated the same experiment. We
% measure an average cost of 33,193.12, 42,465.23 and 23,798.35 gas
% for the $\transfer$, $\approve$ and $\transferFrom$ operations, respectively.
% Thus, our accumulator-based construction imposes a three order of magnitude
% overhead. Furthermore, our construction's operations consume a large portion
% of the block's gas limit which is, currently, about 8 million gas
% (\cite{gaslimit}). The large discrepancy between the gas cost of the
% two constructions' operations, as well as, the small and static
% gas cost of the bare-bones ERC20 token operations, are a by-product of
% Ethereum's flat cost model. The fact that storage-related operations are
% underpriced ((\cite{transactionfeeeco})) and that contracts do not pay a 
% recurring fee proportional to the size of their state is one of Ethereum's
% main incentive misalignments. This issue, if left unchecked, will have severe
% consequences to the security and longevity of, not only Ethereum, but any
% smart contract platform that employs a flat cost model. 
% In Section~\ref{sec:costmodel}, we elaborate more and propose modifications to
% Ethereum's cost model to deal with this issue. In the following, we discuss a
% series of improvements that will diminish the cost of the accumulator-based
% ERC20 token operations.

The security property of the accumulator of Camacho et al. \cite{hashacc} is
based on the presupposition that, prior to an invocation of $\CheckUpdate$ for
the deletion or addition of some element $x$, $x \in X$ or $x \notin X$,
respectively. Thus, prior to, e.g., verifying the addition of some element 
$x$ via the $\CheckUpdate$ algorithm, we have to make sure, via a non
membership witness verification, that $x \notin X$. Part of an ongoing project
is to provide a proof extension that will allow us to lift this assumption
from the accumulator's security property. Consequently, we will be able to
eliminate one, two and four invocations of the accumulator's 
$\Belongs$ algorithm from the $\approve$,
$\transfer$ and $\transferFrom$ operations of the accumulator-based ERC20
token, respectively. Note that one invocation of $\Belongs$ costs, on average, 289,873.23
gas, when $|X| = 400,000$ and will, thus, provide a substantial improvement.

Our implementation of the hash tree accumulator employs the SHA-256 hash
function, which is exposed as a precompiled contract in Ethereum. Precompiled
contracts reside on well-known, static addresses and constitute Ethereum's
``standard library'', similar to that of common programming languages. The
advantage of precompiled contracts is that their computation incurs low
gas costs because their code runs on the miner's machine language. The 
computational cost of the SHA-256 hash function is 60 gas, plus 12 gas per
input word (rounded up) and its implementation complies to the NIST standard.
However, the KECCAK-256 hash function, whose computational cost is 30 gas,
plus 6 gas per input word (rounded up), does not comply to the NIST standard
and is, instead, implemented as an EVM opcode. 
Moreover, precompiled contracts, at each invocation, incur the extra gas cost of a
message call, which is 700 gas. However, that is not the case for EVM opcodes. 
Thus, Ethereum
promotes the use of a non-standard compliant hash function. Recently, a
proposal has been submitted (\cite{precompiledcall}) that suggests the removal
of the message call gas cost for precompiled contracts, which we believe is
fair. Furthermore, we believe that the gas cost of these hash functions should
be equalized. Assuming that both of the aforementioned suggestions are
applied, the gas cost of the hashing operations will be reduced by $93.69\%$
and, as a result, will further diminish the gas cost of the accumulator-based
ERC20 token operations.

To illustrate the overhead of our
accumulator-based ERC20 token construction, we implemented a ``bare-bones''
ERC20 token (where account balances and allowances are stored in the contract's state \cite{sampleerc20}) and repeated the same experiment. We
measure an average cost of 33,193.12, 42,465.23 and 23,798.35 gas
for the $\transfer$, $\approve$ and $\transferFrom$ operations, respectively.
Thus, our accumulator-based construction
is much more expensive, despite its constant and minimal space overhead on
Ethereum's state. The large discrepancy between the gas cost of the
two constructions' operations, as well as, the small and static
gas cost of the bare-bones ERC20 token operations, are a by-product of
Ethereum's flat cost model. The fact that storage-related operations are
underpriced (\cite{transactionfeeeco}) and that contracts do not pay a 
recurring fee proportional to the size of their state is one of Ethereum's
main incentive misalignments. This issue, if left unchecked, will have severe
consequences to the future of, not only Ethereum, but any
smart contract platform that employs a flat cost model. 
Next, we propose modifications to
Ethereum's cost model to deal with this issue.

\section{Revisiting Ethereum's Storage Cost Model}
\label{sec:costmodel}

Ethereum employs a flat cost model to price all EVM opcodes
(\cite{ethyellowpaper}), including storage-related operations.
There are two main issues with this approach. 
First, storing data on the
state of smart contracts incurs a \emph{one time fee} which is
underpriced (\cite{transactionfeeeco}). To our knowledge, there
is no other, real world system that provides such high levels of data replication
and availability without a recurring fee that is proportional to the volume of the stored data. Furthermore, as there is no incentive for freeing storage, Ethereum is
faced with a tragedy of the commons problem with regards to the
monotonically increasing size of its state.
Second, Ethereum's flat cost model does not account for the complexity
of executing storage-related operations, which is a function of the size
of the state of smart contracts.
We propose the following modifications to Ethereum's pricing of storage
to address these issues.

\textbf{Recurring Storage Fees:} The concept of introducing ``storage rent'', i.e., a recurring fee that smart
contracts have to pay based on the amount of storage they consume has been
discussed over the years.
Buterin's original proposal (\cite{rentinit})
has spurred a lot of discussion and has led to the publication of several articles
(e.g.,~\cite{rentarticle1,rentarticle2,rentnow}) which, in their vast majority,
stress how important such a mechanism is for the longevity of public blockchains.
An additional use of the rent mechanism is to clean up Ethereum's state from
accounts (contracts are accounts as well) that are not being used anymore.

Our proposal on the subject of storage rent is based on the following points.
First, we believe that rent fees should not be rewarded to anyone
as that could introduce new attack vectors. Second, since Ethereum
is a global computer, it is rational to assume that it has a predefined
storage capacity $S_{max}$ (e.g., Buterin has suggested 500 GB \cite{rentdisk}). Naturally, this is a conceptual upper bound on
the state's size and will, essentially, reflect an estimate of what is considered reasonable for the
average miner.
Third, $S_{max}$ should be adjustable by the ones that
maintain the network, i.e., the miners, to account for real world, storage trends. This could be achieved via a mechanism similar to
the one that is already in use for adjusting block difficulty. We propose that up to a low utilization percentage of the system's
state, e.g., $U_{low} = 25\%$, the rent per storage key of a contract's
state should be static to reflect
the low burden imposed on miners. When the state's utilization is between $U_{low}$ and, e.g., $U_{high}=80\%$, the rent per storage key of a
contract's state should increase logarithmically with the total number of
keys in the system's state. This reflects the fact that Ethereum's state is organized on top of LevelDB whose complexity we elaborate more on the
following paragraph. From thereon in, rent fees
should be prohibitive, thus, they should scale linearly to the total number of keys in the system's state. To derive a
base rent fee per storage key, we considered real world examples of systems
that are highly replicated, available and charge for storage. 
Cloud storage providers are a prime example. For instance, Amazon's EFS
(\cite{efs}) charges 0.30 USD per GB per month. At the time of this 
writing, one unit of Ether corresponds to 202.18 USD (\cite{usdtoeth}). Based on this analogy,
we compute a base rent fee of $R_{base} = 530,657,634.8$ Wei per storage key per year
(1 Ether corresponds to $10^{18}$ Wei). Thus, we have an adaptable scheme for computing rent fees that follows the 
laws of supply and demand by considering the state's overall utilization and the burden imposed on miners.

\textbf{Scaling Storage Costs:} A contract's state is organized
on an on-disk Merkle Patricia (MP) trie (\cite{patriciatree}), which is referred to
as the \emph{storage trie}. This is a modified version of a
typical radix tree with the added property of Merkle trees, i.e., the root hash
uniquely identifies the (key,value) pairs in the tree. The nodes of the
storage trie and the smart contract's state (storage keys) are stored in a
LevelDB (\cite{leveldb}) key-value store, whose underlying data
structure is a multi-level Log Structured Merge (LSM) tree.
As illustrated in a recent study (\cite{mlsm}), due to
Ethereum's authenticated storage (MP trie), one Ethereum read (e.g.,
reading the root node of a contract's storage trie) can lead to 64 LevelDB 
$\mathsf{get()}$
(read) requests. Each $\mathsf{get()}$ may internally involve multiple disk
reads due to the large amount of metadata that LevelDB maintains (\cite{lsmratio}). Updates to a contract's storage, e.g., adding/updating
storage keys, result in updates to its storage trie that have to be
committed on disk. In LevelDB, key-value updates are reinserted into a skip list with a monotonically increasing sequence
number along with a ``tombstone'' flag that invalidates the pair's prior
version. To maintain key-value pairs in sorted order, LevelDB uses
a compaction method. This process involves multiple merge sorts (one per 
LSM tree level)
and incurs
a write amplification factor, which is the ratio of the amount of data written to the amount of data requested for writing by users,
of $\times11$ (\cite{lsmratio}). 

\begin{figure*}
    \centering
    \begin{subfigure}{0.3\textwidth}
        \includegraphics[width=.78\linewidth,angle=270]{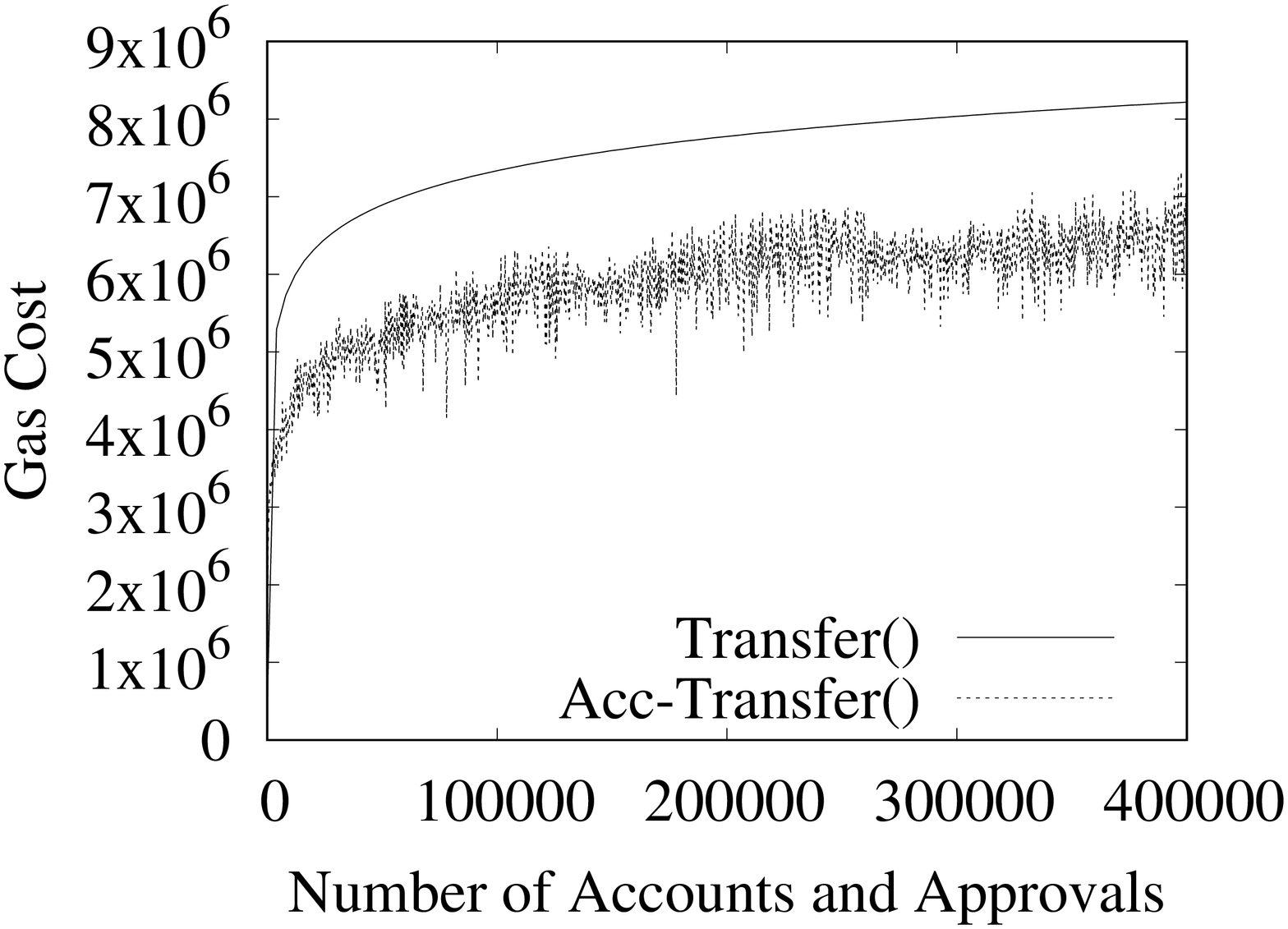} 
        \caption{}
        \label{subfig:transfer}
    \end{subfigure}
    ~
    \begin{subfigure}{0.3\textwidth}
        \includegraphics[width=.78\linewidth,angle=270]{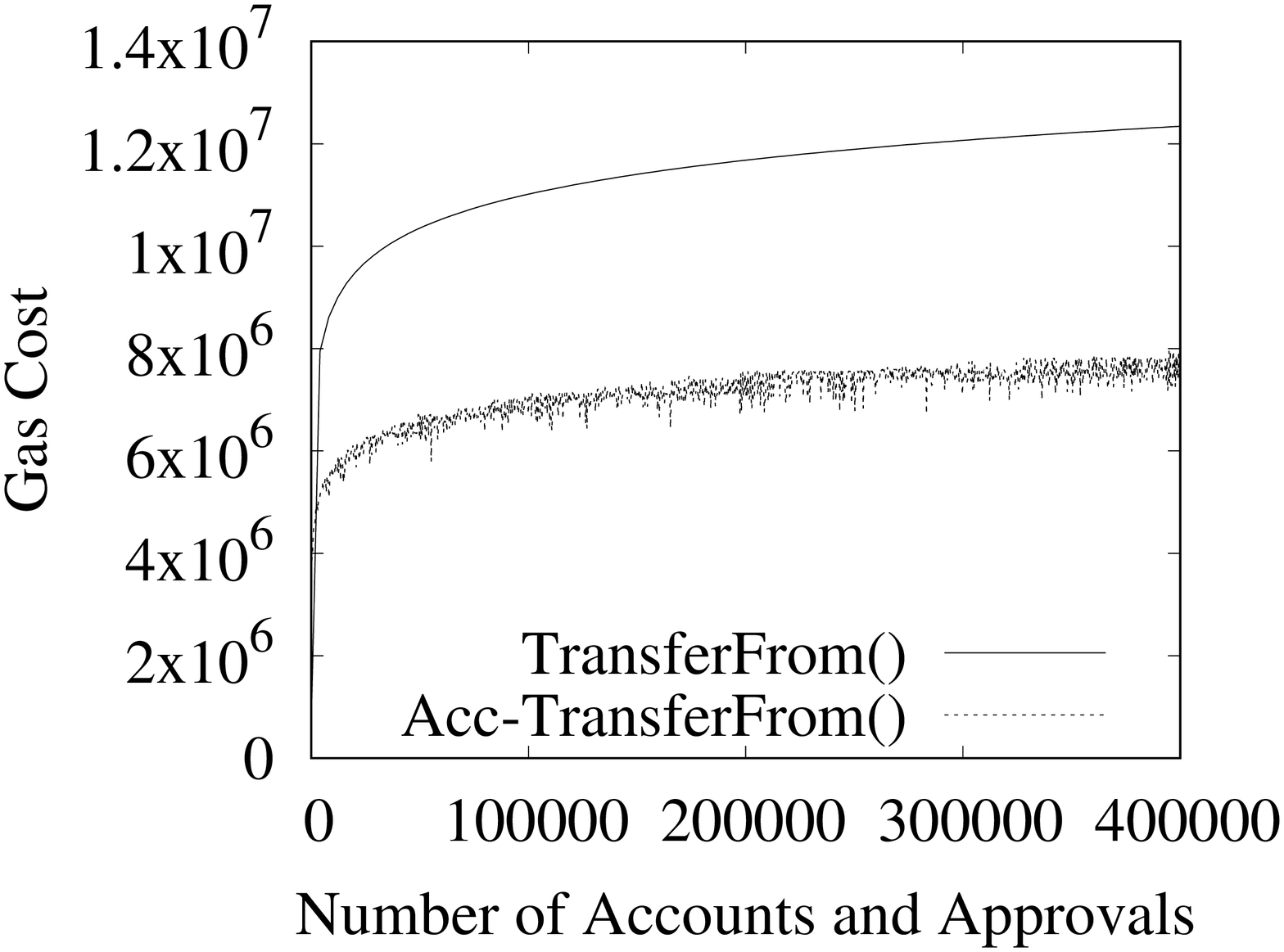}
        \caption{}
        \label{subfig:transferfrom}
    \end{subfigure}
    ~
    \begin{subfigure}{0.3\textwidth}
        \includegraphics[width=.78\linewidth,angle=270]{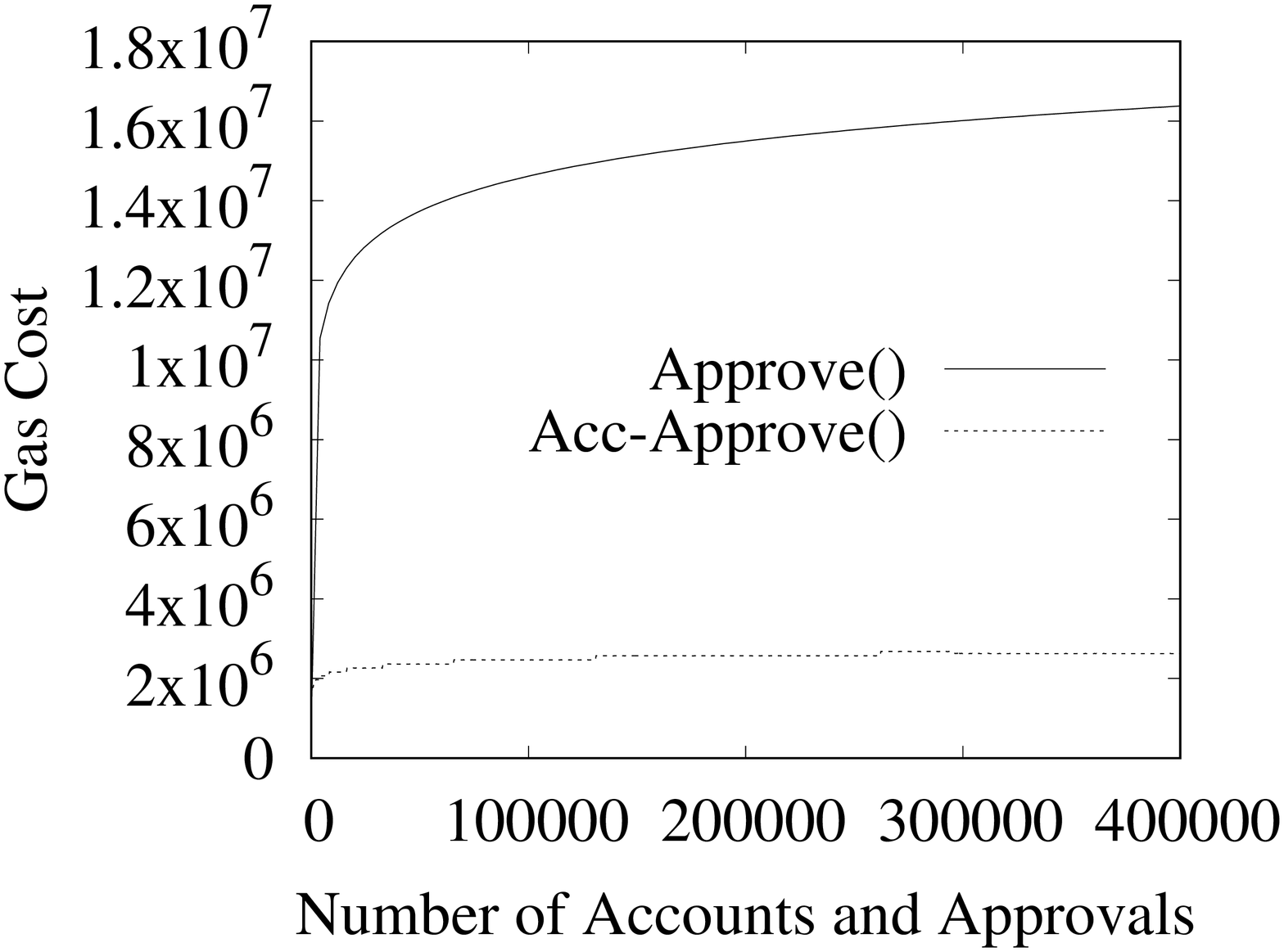}
        \caption{}
        \label{subfig:approve}
    \end{subfigure}
\caption{Gas cost versus of the 
$\transfer$ (\subref{subfig:transfer}), $\transferFrom$ (\subref{subfig:transferfrom}) and $\approve$ (\subref{subfig:approve}) operations of the ERC20
token and the accumulator-based ERC20 token for up to a total of 400,000
accounts and 400,000 approvals under the new storage cost model.}
\label{fig:costmodel}
\end{figure*}

Ethereum's flat cost model does not reflect the aforementioned complexity of storage-related operations. One might assume that an ideal scheme would scale the cost of
these operations based on the number of incurred disk operations. However,
this is not possible as Ethereum miners do not have a shared hardware
configuration, e.g., their physical hard disks and their caches vary
significantly. This would interfere with Ethereum's consensus as the
execution of the same transaction would lead to different gas costs across
different miners. 
% Furthermore, miners are, largely, agnostic to LevelDB's
% internals, e.g., LevelDB does not expose details regarding the number of reads
% and writes during its compaction process. 
Instead, we
propose a scheme where the cost of storage-related operations is computed on
a per transaction basis and scales according to the number of operations to
LevelDB's LSM tree, which is the same across all miners. Fetching one key from a LSM tree involves two binary searches (\cite{pebbles}). Accessing the value of a smart contract's storage 
key involves, at minimum, fetching one node of its storage trie and,
subsequently, fetching the storage key itself. Thus, it requires a total of
four binary searches, i.e., $4\log_{2}(n)$ accesses, where $n$ is the number
of storage keys. Updating, or, adding a new storage key, involves the
same number of accesses to infer the value of the tombstone flag. However, since updates are propagated to all levels of LevelDB's LSM tree
during its compaction process, they are subject to LevelDB's write amplification factor, which we discussed above. Thus, updates incur a total of
$11\times4\log_{2}(n)=44\log_{2}(n)$ operations.
Currently, reading, storing and updating storage keys costs 200, 20,000 and
5,000 gas, respectively. Thus, under our proposed scheme, the cost
of, e.g., reading a storage key is $200\times4\log_{2}(n)$.

Figure~\ref{fig:costmodel} illustrates the gas cost
of
the $\transfer$, $\transferFrom$  and $\approve$ operations of the bare-bones
and our accumulator-based ERC20 token
under our proposed cost model. Regarding the bare-bones ERC20
token, we only plot the storage-related cost of its operations, which are the
dominant factor. The biggest discrepancy is in the
$\approve$ operation (Figure~\ref{subfig:approve}) where our accumulator-based
construction provides an order of magnitude improvement. Overall, results illustrate that, under a pricing scheme that reflects
the effort that miners have to expend to execute storage-related operations,
the programming paradigm that we propose in this work provides reduced gas 
costs across all ERC20 token operations. Nevertheless, we believe that the most
important property of our approach is that it aligns well with the future
of smart contract platforms since it incurs constant storage
overhead to miners.

\section{Conclusion}
\label{sec:conclusion}
% We propose a scheme for computing rent fees that follows the
% laws of supply and demand by considering the state's overall utilization and the burden imposed on miners. In addition, our scheme is adjustable
% to real world, storage trends. Furthermore, we introduce scaling of the
% cost of storage-related operations to account for the effort that miners
% have to expend to execute them. Lastly, we introduce an alternative
% programming paradigm for developing dApps
% that promotes diversity, scalability and aligns well with the
% future of smart contract platforms.
We introduce an alternative
programming paradigm for developing dApps
that promotes diversity, scalability and aligns well with the
future of smart contract platforms. Our approach can be adapted to
any application that requires a verifiable representation of its application data. 
We propose a scheme for computing rent fees that follows the
laws of supply and demand by considering the state's overall utilization, as well as the burden imposed on miners. In addition, our scheme is adjustable
to real world, storage trends. We introduce scaling of the
cost of storage-related operations to account for the effort that miners
have to expend to execute them. Lastly, we show
that under such a pricing scheme that encourages economy in the state consumed by
smart contracts, our ERC20 token adaptation reduces the incurred
transaction fees by up to an order of magnitude.

\bibliographystyle{IEEEtran}
\bibliography{IEEEabrv,bibliography}
\end{document}